\documentstyle[prb,aps,multicol]{revtex}
\begin{document}
\draft

\title{\bf Charge redistribution at Pd surfaces:\\
	{\it ab initio} grounds for {\it tight-binding} interatomic potentials }

\author {S. Sawaya, J. Goniakowski, C. Mottet, A. Sa\'ul and G. Tr\'eglia}
\address {CRMC2, Centre National de la Recherche Scientifique$^{*}$,
Campus de Luminy, Case 913, 13288 Marseille CEDEX 9, France.}

\date{\today}

\maketitle

\begin{abstract}

A simplified {\it tight-binding} description of the electronic structure
is often necessary for complex studies of surfaces of transition metal compounds.
This requires a self-consistent parametrization of the charge redistribution,
which is not obvious for late transition series elements (such as $Pd$, $Cu$, $Au$),
for which not only $d$ but also $s-p$ electrons have to be taken into account.
We show here, with the help of {\it ab initio} FP-LMTO approach,
that for these elements the electronic charge is unchanged from bulk to the surface,
not only per site but also per orbital. This implies different level shifts
for each orbital in order to achieve this {\bf orbital neutrality} rule.
Our results invalidate any neutrality rule which would allow charge
redistribution between orbitals to ensure a common {\bf rigid shift} for all of them. 
Moreover, in the case of $Pd$, the power law which governs the 
variation of band energy with respect to coordination number,
is found to differ significantly from the usual {\it tight-binding} square root.

\end{abstract}

\begin{multicols}{2}
\section{Introduction}
\noindent

Transition metal materials are widely studied due to their remarkable properties
in various domains such as metallurgy, electronics, magnetism or catalysis.
These properties can be related to the particular electronic structure of transition
metals, characterized principally by the progressive filling (up to ten 
electrons) of a narrow $d$ band. However, when this band becomes nearly filled
(i.e. for elements at the end of the transition series)
the role of external $s$ and $p$ electrons can no longer be neglected.
Therefore, any relevant theoretical treatment requires to account
simultaneously for the two different types of electronic states:
$s$ and $p$ electrons with strong itinerant character, and $d$ electrons with 
partially localized one.
Such treatments involve, depending on the problem, two different types of 
methods, of opposite complexity and flexibility.
On the one hand {\it semi-empirical} methods, which require fitted parameters
	but can be applied to large systems (a few thousands of atoms)
	with complex atomic (defects, surfaces) or chemical (alloys) structure.
On the other hand {\it ab initio} methods, which do not need any parameters
	but are limited to small systems of only a few tenths of atoms.
Now, the modern challenge in material science is to predict both the atomic
and chemical structure of complex systems (such as alloy surfaces) 
from the only knowledge of the electronic structure of their components. 
This requires large scale kinetic simulations of the Monte Carlo or 
Molecular Dynamics type, in the framework of interatomic potentials 
which have to be as simple as possible (i.e. preferably in analytical form) 
while remaining fine enough to describe correctly differences in metallic 
cohesion of elements.
In particular, one has to go beyond simple pair interaction potentials,
which have proven to be inadequate for transition metals.
Appropriate potentials can be obtained from semi-empirical electronic
structure calculations, suited to the character of the electrons involved:
{\it effective medium theory} for $s-p$ electrons \cite{nor80}, 
{\it embedded atom model} \cite{foi86}
or {\it tight-binding} approximation \cite{fri69} for $d$ electrons. 
More precisely, in the framework of {\it tight-binding} approximation,
different types of potentials have been derived,
depending on the type of addressed problem: {\it second moment many-body} potentials
for atomic relaxations \cite{ros89} or {\it effective pair-wise} potentials
for ordering processes in alloys \cite{tre88}.
However, it is worth noticing that these potentials have been obtained
under drastic assumptions concerning the charge redistribution close to
the surface (or more generally, close to a defect). 
In particular, the commonly admitted rule is that of local charge neutrality 
at the defect. 
Let us emphasize that infringing this law can change completely the 
essential features of the simplified models mentioned above. 
It is thus essential, to validate the semi-empirical potentials in use, 
or to propose alternative ones, starting from some parameter--free 
reference calculations and deriving general trends for the variation of 
charge transfer near the defects as a function of coordination and of $d$ band 
filling.

Beyond these applications which only require an overall knowledge of the 
local density of states since they involve integrated quantities 
(energy, band filling), a precise knowledge of the local densities is 
needed for direct comparison with experiments (angular photoemission) 
or for properties related to details of the density of states near the 
Fermi level (reactivity, magnetism). 
The {\it tight-binding} formalism is in general sufficient to account for these
requirements as long as $d$ electrons play the predominant role,
i.e. for not too large $d$ band filling\cite{fri69}. 
The problem becomes complicated for elements belonging to the end of transition series 
(which are the most commonly used for their catalytical ($Pd$) or magnetic
($Ni$) properties). The $d$ band of these elements is nearly filled
so that the role of $s-p$ electrons can no longer be neglected.
But an unified description of $s-p$ and $d$ electrons is tricky since
the strong itinerant character of the former deserves nearly free 
electron approaches while the less delocalized character of the 
latter is the basis of the {\it tight-binding} approximation.
In spite of this apparent incompatibility, it is commonly admitted that 
one can take into account the $s-p$ and $d$ hybridization in the 
framework of the {\it tight-binding} approximation \cite{mot96}. 
However, this implies to extend the local neutrality rules near the defects, 
previously used for $d$ orbitals only, to the case where orbitals of different
character are involved.  
It is then essential, before using extensively this type of approach, to 
analyze the assumptions in the light of {\it ab initio} calculations in 
which both $s-p$ and $d$ electrons are treated on the same footing.

We touch here an aspect of a more general problem of transferability
of tight-binding parameters to a new environment and of conditions these 
parameters should meet in order to describe the electronic structure of 
modified atomic and chemical configurations which can occur in actual 
materials.
More precisely, they should remain valid in a wide range of atomic 
coordinations and environments, apply over a range of interatomic separations
and describe accurately the chemical effects between unlike species.
It is thus essential to examine them in the light of {\it ab initio} type 
calculations, considered as exact reference. Several attempts have already 
been made to check the reliability and usefulness of these tight-binding 
parameters by examining their transferability with changes in 
structure\cite{rob91,met92,fab94} and chemical environment \cite{slu94}.
However, at present only few studies can be found on what concerns the 
influence of the modified surface environment on the $sp-d$ hybridization
and induced $sp-d$ charge transfers.

Therefore, the aim of the present paper is to review, following a similar approach,
the alternative hypothesis underlying the semi-empirical methods as applied
to transition metals of the end of the series and to validate some simple rules
to be used in {\it tight-binding} approaches.
In order to do this we analyze the variation of the local electronic structure 
(site projected densities of states, level shifts, charge transfer) from bulk to low
index surfaces of $Pd$ obtained within a parameter-free, FP-LMTO
approach. We then derive the general trends and use them as to modify 
accordingly the {\it tight-binding} framework. 
The paper is organized as follows: in the Section II we present to some
extent the principles of the {\it tight-binding} approach and the question marks
on the parametrization one has to answer. Section III is devoted to a brief
description of the FP-LMTO method which we have chosen for the reference 
calculations.
In Section IV we outline our {\it ab initio} results and compare them to
results obtained within {\it tight-binding} approach based on alternative 
neutrality rules. We discuss their respective similarities and differences 
as well as general tendencies which can be due to application of this or the other 
assumption.

\section{Tight-Binding treatment}
\noindent
We will use in the following the usual {\it tight-binding} hamiltonian 
extended to the case of multiple orbitals :
\begin{equation}
H=\sum_{i,\lambda} |i,\lambda>  \epsilon_{i\lambda}^0 <i,\lambda| + 
\sum_{\lambda\mu,\ i\neq j} |i,\lambda>\beta_{ij}^{\lambda\mu}<j,\mu|
\label{equ:HTB}
\end{equation}
It involves two types of parameters:
\begin{itemize}
  \item $\epsilon_{i\lambda}^0$=$\epsilon_\lambda^0$+$\delta\epsilon_{i\lambda}$:
	where $\epsilon_\lambda^0$ is the atomic level for the orbital 
        $\lambda$ ($s$, $p$, $d$),
	and $\delta\epsilon_{i\lambda}$ is the shift of the atomic level for orbital
	$\lambda$ at site $i$, which is required to achieve the charge self-consistency
	on any site $i$ which is not equivalent to a bulk site (surface, defect) \cite{spa85}.
  \item $\beta_{ij}^{\lambda\mu}$: hopping integral between the orbital
        $\lambda$ at site $i$ and the orbital $\mu$ at site $j$
\end{itemize}
These hopping integrals can be expressed in terms of the usual Slater parameters.
They will be derived here from the band structure using the interpolation scheme
developed by Papaconstantopoulos \cite{pap86}.

The local densities of states $n_{i\lambda}(E,\delta\epsilon_{i\lambda})$
 are obtained from the
continued fraction expansion of the Green function $G(E)=(E-H)^{-1}$,
the coefficients of which are directly related to the
first moments of the density of states,
and calculated within the {\it recursion} method\cite{hay75}. 
It is then possible to define the band filling per orbital $N_{i\lambda}$,
which is obtained by integration of the partial local density of states
up to the Fermi level $E_F$:
\begin{equation}
N_{i\lambda}=\int_{-\infty}^{E_F} 
{n_{i\lambda}(E,\delta\epsilon_{i\lambda})dE}.
\label{equ:remp1}
\end{equation}
Total band filling (number of valence electrons) can be obtained by
summing over all orbitals $N_i=\sum_{\lambda} N_{i\lambda}$.

The problem is then to calculate the level shift $\delta\epsilon_{i\lambda}$
corresponding to the charge self-consistency on defect sites.
When only the $d$ electrons need to be taken into account
(elements in the middle of transition series),
it has been shown\cite{all76} that the charge self-consistency 
reduces to a {\it local charge neutrality condition}. According to this
condition, number of $d$ electrons on a defect site (surface) is the same 
as on the bulk site. In other words, whatever the site $i$, 
$N_{id}=N_{bulk,d}$.
Then, all the inner levels are assumed to follow rigidly this valence $d$-band shift,
giving rise to surface core level shifts
evidenced by photoemission experiments\cite{spa85}.
The situation is more intricate when orbitals of different character have to 
be taken into account.  Actually, one can then wonder if a similar local 
neutrality rule still applies ($N_i=N_{bulk}$), and if so has it to be achieved
for each of orbitals separately ($N_{i\lambda}=N_{bulk,\lambda}$)
or not ($N_{i\lambda} \neq N_{bulk,\lambda}$)?
\begin{itemize}
  \item If it is true ($N_{i\lambda}=N_{bulk,\lambda}$),
	one should then introduce three different level shifts
	($\delta\epsilon_{is} \neq \delta\epsilon_{ip} \neq \delta\epsilon_{id}$)
	to ensure each orbital neutrality.
  \item If it is not true, i.e. that a charge redistribution occurs between
	the different orbitals ($N_{i\lambda} \neq N_{bulk,\lambda}$),
	then the neutrality rule contains too much degrees of freedom,
	and we have to introduce additional assumptions on the shifts.
	A usual one is then to follow the rule for inner shell
	and to impose a common {\bf rigid} shift for all orbitals\cite{mot96}
	($\delta\epsilon_{is} = \delta\epsilon_{ip} = \delta\epsilon_{id}$).
\end{itemize}

The choice is far from being harmless as will be seen in section IV. 
If one remembers that these are quantities which are closely related to
reactivity (adsorption energies) or magnetic properties, one has to 
be able to decide which approximation approaches better the reality.
One solution would be to compare these results to experimental data,
for instance to the values of the core level shifts. However, this is 
not completely unambiguous, since the respective influence of initial and 
final state effects cannot be precisely estimated. 
The best way to proceed is then to perform some {\it ab initio} calculation 
leading to the full charge distribution, and to analyze how it can be shared 
between different sites and different types of orbitals.

\section{LMTO method}
\noindent
To perform these reference parameter--free calculations we have chosen 
a LDA based FP-LMTO method\cite{lmto,lmto1}. It has proven to be well suited for
the description of bulk electronic structure of transition metal 
compounds\cite{trans,trans1}, and no approximation on shape of charge density nor 
potential makes it also adequate for low symmetry systems such as 
surfaces\cite{sur,sur1}.

Within the FP-LMTO method the space is divided into non-overlapping spheres
centered on atomic sites. The basis set consists of atom-centered Hankel
envelope functions which are augmented inside atomic spheres by means of
numerical solution of scalar-relativistic Dirac equation. Due to the 
non vanishing interstitial region it is enough to use the minimal basis set - 
we have used three $s$, three $p$ and three $d$ functions per atomic site
(for three different energies: -0.7, -1.0 and -2.3~Ry) \cite{sur} corresponding to 
three different localization of Hankel envelopes.
We have used 'two-panel' technique as to include $4p$ semicore electrons
as full band states. Valence band contains $4d$, $5s$ and $5p$ states.

To obtain an accurate representation of the exponentially decaying density
outside the surface, in slab calculations it is necessary to cover the 
surfaces with one or two layers of empty spheres. We have included spheres 
which are the first neighbors of surface metal atoms. One layer of empty 
spheres was thus used on (111) and (100) faces and two layers on the
more open (110) face. The empty-sphere angular-momentum cutoff for
charge density and for the augmentation of the wave function was fixed to 6 and 4, respectively.

We have assured ourself that the basic experimental bulk characteristics of
palladium are reproduced correctly (with respect to measured values: lattice 
parameter -1\%, bulk modulus 10\%, cohesive energy 20\% - overestimation 
of the latter is however one of well known deficiencies of the LDA).
The lattice constant determined for the bulk was systematically used for 
slab calculations. Surface relaxation effects were not taken into account.
We have tested the convergence of results with respect to the slab thickness
(in order to obtain bulk-like properties in the center of the slab) using
as a criterion the modifications of calculated surface energy. Increasing the 
number of layers from 9 to 11 changes the latter by less than 40 meV per surface atom.
Nine layer slabs were thus adopted for the calculations.

The {\bf k} point sampling of the Brillouin zone was done with special point 
meshes and converged to within 50 meV/atom (with the Gaussian broadening of 
20 mRy) for 126 {\bf k} points in the irreducible part of BZ for (100) surface, 216 {\bf k} points
for (110) and 226 {\bf k} for (111). 

\section{Results and Discussion}
\noindent

Before presenting our results, it is worth noticing that we have performed
calculations for various FCC metals belonging either to the end of the
transition metal series ($Pd$, $Ni$) or to the noble metal column
($Cu$, $Ag$, $Au$) for which the role of $s-p$ electrons is yet
much predominant. Since the essential results are the same
for all these elements, we will only present here the case of $Pd$,
which is likely the element which has been the most widely studied
in relation with its unique catalytical properties.
In order to use the conclusions of the {\it ab initio} calculations as a ground
for the {\it tight-binding} description of surfaces we need to start by checking 
if both methods lead to similar results for the bulk electronic structure.
Having a look at Figure 1, where we have plotted the bulk LDOS calculated 
with the two approaches one can see that the structure of {\it tight-binding} 
bulk LDOS is in good agreement with the corresponding {\it ab initio} results,
provided that the former one is calculated with a sufficiently large number
of exact moments. Here a number of eighty has been used in order to
reproduce all the details of the {\it ab initio} LDOS. 
The comparison between the dotted curves in Figure 1 show that the interpolation scheme
is correct and that, at least in the bulk, {\it tight-binding} formalism
can be successfully applied to describe elements 
of the end of the transition series, by taking into account not only valence 
$d$ states but also $s-p$ ones.

The situation is obviously more complex at a surface. As explained in 
Section II, in order to propose a {\it tight-binding} self-consistent calculation
scheme, one has to know how the electronic charge distribution on the surface
differs from that in the bulk. Indeed, the knowledge of the charge transfers
is required to write the charge self-consistency rule which will then allow us to
determine, using an iterative scheme, the level shifts: $\delta\epsilon_{i\lambda}$.
We have thus analyzed the spacial redistribution of electrons due to surface creation, 
as obtained within the FP-LMTO approach. 
In Figure 2 we plot, for two surface orientations ( (111) and (100) ), a contour map
of the difference between valence electron densities obtained respectively for slab
and bulk geometries, cross-sectioned along a plane perpendicular to the surface.
It can easily be seen that the  overall modifications are very small. They are
principally constrained to within the surface atomic layer and consist of polarization
of surface atoms related to depopulation of orbitals pointing into vacuum.
On the other hand, some electrons are spilled out into the vacuum region 
corresponding to hollow surface sites.

In order to quantify these effects better, we give in Table I the charge
distributions calculated inside ($s$, $p$ and $d$ orbitals) and
outside (interstitial region and hollow surface sites) the atomic MT spheres,
for atoms on surfaces of three inequivalent orientations and,
for the sake of comparison, in a bulk plane.
The electronic occupation for atoms in the first underlayers are not given
since they are found to be quasi-identical to those in the bulk.
Separation into the two regions (inside and outside the spheres)
avoids any arbitrary reattribution of interstitial electrons
to particular atomic sites.
However, when trying to reproduce the above charge distribution with the
{\it tight-binding} approach, we will have to remember that its basis set is
not conceived as to represent off-site charges.
The only solution will then be to reattribute electrons found
in the hollow surface sites to some on-site atomic orbitals. 
Due to spacial asymmetry and to their relative delocalization with respect 
to surface atoms, these are the $p$ orbitals that seem to be the most 
adequate candidates. This means that, for the $p$ charge distribution,
we would add the charge found in the hollow sites to that of the $p$-orbital.
Once the electrons reassigned in this way, one can see that the differences
between surface and bulk distributions are indeed very small.
The electron transfer is even less pronounced than in the contour plot,
because it is integrated over orbitals of the same character:
on-site electron transfer between $d$ orbitals of different orientations,
which can be seen in the plot, does not influence $s$, $p$ and $d$ - projected
electron occupations.

As a conclusion, one can say that the self-consistent treatment of the charge 
at a surface of a system with $sp$-$d$ hybridization can be reduced to a 
condition of local charge neutrality {\bf per orbital}.
It is worth noting that this condition is not compatible with the rigid shift
$\delta\epsilon_i$ used in old scheme\cite{mot96}.
 
Let us now see to what extent a {\it tight-binding} calculation,
based on this condition of charge neutrality per orbital,
indeed reproduces correctly the {\it ab initio} results.
To this aim, we first compare, in Figure 1, the LDOS for (111) and (100) surface sites
calculated by either the {\it tight-binding} approach (Fig.1b-1b')
or the FP-LMTO method (Fig.1a-1a').
One can see that the overall deformation induced by the surface
is well reproduced in this framework.
In order to make the comparison more quantitative we have looked on basic
characteristics of surface bands. In particular we have compared the first moment of
LDOS, related directly to the surface level shift.
As can be seen in Table II, both {\it ab initio} and {\it tight-binding}
calculations predict an upward surface level shift,
smaller for compact (111) surface and enhanced for the most open (110) one.
The values are found in good agreement:
they should now deserve some experimental confirmation from core level 
spectroscopy or from surface adsorption energies.

In order to illustrate how crucial it is to achieve properly this
charge self-consistency, we have also performed the {\it tight-binding} calculation
by assuming, as in some previous works\cite{mot96},
that at the surface the charge can be redistributed between the different orbitals,
and that the latter undergo the same rigid level shift.
Comparing the LDOS obtained in this way (see Fig.1c-1c')
to the previous ones (see Fig.1b-1b') is meaningful.
Indeed, the detailed structure and position of peaks in the LDOS are different, 
in particular around the Fermi level and in the bottom of the band,
where the $sp$-$d$ hybridization is the most important.
This could have been expected since the structure of the bottom of the band is very
dependent on the respective positions of the levels for $s$, $p$ and $d$ orbitals.
The comparison between these curves and the {\it ab initio} ones (see Fig.1a-1a')
shows a significantly better agreement for those obtained by assuming orbital
charge neutrality than a common rigid shift.
This conclusion can be made more convincing by comparing the surface level shift.
Indeed, as can be seen in Table II, one still recovers,
within the rigid shift assumption, an upward surface level shift,
with the same trend from the compact to the open surface,
but with a significantly too low value,
which can be attributed to an over-estimated electron transfer
from $s$ and $p$ to $d$ orbitals on sites of reduced coordination.

One can then conclude that the fundamental difference induced
by alternative neutrality conditions underlying the {\it tight-binding} approach
is the way the $s$ and $p$ electrons are transferred to $d$ orbital when the 
coordination number decreases. Although in the atomic limit of $Pd$ all 
electrons are to be found  in the $d$ shell, the approach which allows
charge redistribution predicts strong transfers already for relatively high 
coordination numbers. Our FP-LMTO results suggest that this is not the case and that
a constant band-filling (orbital neutrality) corresponds better to calculated trends.
However, the limits of applicability of the orbital neutrality condition
for very low coordination number are not still unambiguously defined.

Finally, it is interesting to compare also the values of the second moments $\mu_2$
(i.e. the bandwidth at half-maximum) of the {\it tight-binding} $d$-LDOS
obtained for the bulk and three different surface orientations ($Z=12,9,8,7$),
to those derived from LMTO calculations.
Indeed, a precise knowledge of the variation of $\mu_2$ with the coordination number
($Z$) is essential when interested in the definition of simple many-body interatomic
potentials to be used in numerical simulations.
In particular, this could give us an alternative to the so-called
{\it second moment} approximation \cite{ros89},
in which the attractive part ($E_b$) of the potential
is fitted to the band energy of a rectangular density of states,
of same second moment as the actual density of states.
This constraint leads to:
\begin{equation}
E_b \sim \mu_2^{1/2}.
\label{equ:eb}
\end{equation}
The problem is then to calculate this second moment.

In the {\it tight-binding} approximation, it is known to be
linear in the effective coordination number:
\begin{equation}
\mu_2^{TB} \sim Z,
\label{equ:mu2tb}
\end{equation}
so that the Eq.(\ref{equ:eb}) can be re-written:
\begin{equation}
E_b^{TB}$ $\sim$ $Z^{1/2},
\label{equ:ebtb}
\end{equation}
which leads to the usual {\it square-root} many-body character
of this potential \cite{ros89}. However, in spite of its overall success,
this potential sometimes fails to reproduce the experimental variation
of the energy as a function of the coordination number in the whole $Z$-range,
in particular for very low values of $Z$.
This is in particular the case for $Pd$ \cite{rou95}.
The alternative is then:
\begin{itemize}
	\item either to go beyond the second moment approximation to account for
		details of the LDOS which are obviously neglected when using a
		rectangular shape,
	\item or to stay within such a second moment approximation, provided that
		one calculates the second moment as accurately as possible,
		going beyond the {\it tight-binding} approximation
		by using {\it ab initio} calculations.
\end{itemize}

The first solution has already been explored, but without $sp-d$ hybridization,
by other authors \cite{gue95}, leading to $E_b^{TB}$ $\sim$ $Z^{2/3}$.
We will then illustrate here the consequences of choosing the second solution. 
In that case, deriving  $\mu_2$ from our LMTO calculations.
Indeed, this leads to:
\begin{equation}
\mu_2^{LMTO} \sim Z^{3/2},
\label{equ:mu2lmto}
\end{equation}
which, used in the eq.(\ref{equ:eb}), gives a somewhat different $Z$-dependence
for the attractive term:
\begin{equation}
E_b^{LMTO}$ $\sim$ $Z^{3/4}.
\label{equ:eblmto}
\end{equation}
This means that a proper treatment of $sp-d$ hybridization
leads to a dependence of the band term which is intermediate between
the square-root ($E_b$ $\sim$ $Z^{1/2}$) and a pairwise ($E_b$ $\sim$ $Z$) dependence.
Actually, this result seems to be confirmed in the particular case of $Pd$,
for which the experimental variation of the energy with coordination number
is found to be almost pairwise.
However, it is worth pointing out that,
contrary to our conclusions concerning charge neutrality,
this result is not general and should be actualized for each different element.

\section{Conclusion}
\noindent
As a conclusion, comparison of calculated characteristics of the electronic
structure obtained within an {\it ab initio} FP-LMTO method and within a 
semi-empirical {\it tight-binding} approach shows that it is still legitimate to 
treat the surfaces of elements of the end of the transition series such as Pd
with the latter.
However one has to take care that the $sp$-$d$ hybridization is correctly 
taken into account. To this aim, one has to achieve the proper
neutrality condition underlying the {\it tight-binding} approach
to compact surfaces. What we learn from the above {\it ab initio} calculations
is that the electronic occupation of orbitals does not depend on the environment. 
Under this assumption, it is therefore possible to apply all the conclusions 
obtained for systems with only $d$ electrons, and in particular to use the 
simplified potentials developed to treat either morphology and dynamics 
({\it second moment approximation}) or surface segregation ({\it tight-binding 
Ising model}).

\end{multicols}

\begin{table}
\begin{tabular}{lcccc}
$ $ &
$Pd(bulk)$ & $Pd(111)$ & $Pd(100)$ & $Pd(110)$ \\
\tableline
$N_{s}$		& $ 0.39$ & $ 0.37$ & $ 0.35$ & $ 0.34$ \\
$N_{p}$		& $ 0.34$ & $ 0.25$ & $ 0.22$ & $ 0.19$ \\
$N_{d}$		& $ 8.09$ & $ 8.12$ & $ 8.14$ & $ 8.16$ \\
$Interstitial$	& $ 1.18$ & $ 1.13$ & $ 1.12$ & $ 1.05$ \\
$Hollow \ sites$&         & $ 0.13$ & $ 0.17$ & $ 0.26$ \\
$Total$		& $10.00$ & $10.00$ & $10.00$ & $10.00$ \\
\end{tabular}
\caption{Orbital projected charge distribution inside the MT sphere ($s$, $p$, $d$ orbitals)
	and outside (either in the interstitial region or in the hollow surface sites),
	calculated within FP-LMTO approach.}
\end{table}

\begin{table}
\begin{tabular}{lccc}
$ $ &
$Pd(111)$ & $Pd(100)$ & $Pd(110)$ \\
\tableline
$FP-LMTO$                        												& 0.38 & 0.54 & 0.63 \\
$T.B.: \ orbital \ neutrality \			(\delta\epsilon_{is} \neq \delta\epsilon_{ip} \neq \delta\epsilon_{id})$	& 0.36 & 0.55 & 0.64 \\
$T.B.: \ orbital \ charge \ redistribution \	(\delta\epsilon_{is}   =  \delta\epsilon_{ip}   =  \delta\epsilon_{id})$	& 0.17 & 0.18 & 0.25 \\
\end{tabular}
\caption{Surface level shift $\delta\epsilon_{id}$(eV) for $d$ orbitals on three
	inequivalent surfaces deduced from $FP-LMTO$ calculations
	and calculated with {\it tight-binding} approach ($T.B.$)
	based on alternative neutrality hypothesis}
\end{table}

\begin{figure}
\caption{$Pd$ bulk (dotted curve) and surface (full curve) local densities of states
	for two orientations: (100) (left hand side) and (111) (right hand side).
	(a)-(a'): FP-LMTO calculation,
	(b)-(b'){\it tight-binding} assuming an orbital charge neutrality rule,
	(c)-(c'){\it tight-binding} assuming orbital charge redistribution
		and a rigid level shift.}
\end{figure}

\begin{figure}
\caption{Contour map of the {\it excess} electronic density 
	$\Delta n(r) = n_s(r) - n_b(r)$. $n_s(r)$ is the actual electronic density 
	calculated for the slabs geometries and $n_b(r)$ the one calculated 
	replacing all the atoms of the slabs with bulk ones.
	The figures represent surface cuts perpendicular to the 
	(100) (left hand side) and (111) (right hand side) surfaces.
	A grey scale proportional to $\Delta n(r)$ was used,
	white corresponds to the minimum negative value and 
	black to the maximum positive one.}
\end{figure}

\end{document}